\documentclass[twocolumn,showpacs,preprintnumbers,amsmath,amssymb,prl]{revtex4}
\usepackage{graphicx}% Include figure files
\usepackage{dcolumn}% Align table columns on decimal point
\usepackage{bm}% bold math

\begin{document}

\title{Turning light into a liquid via atomic coherence}
\author{Humberto Michinel and Mar\'ia J. Paz-Alonso}
\affiliation{\'Area de \'Optica, Facultade de Ciencias de Ourense,\\ 
Universidade de Vigo, As Lagoas s/n, Ourense, 32004 Spain.}
\author{V\'ictor M. P\'erez-Garc\'ia}
\affiliation{Departamento de Matem\'aticas, E.T.S.I. Industriales,\\ 
Universidad de Castilla-La Mancha, Ciudad Real, 13071 Spain.}

\begin{abstract}
We study a four level atomic system with electromagnetically induced transparency with giant $\chi^{(3)}$ and $\chi^{(5)}$  susceptibilities of opposite signs. This system would allow to obtain multidimensional solitons and  light condensates with surface tension properties analogous to those of usual liquids. 
\end{abstract}


\maketitle
%------------------- INTRODUCTION -----------------------------------------
The applications of nonlinear optical media mostly rely 
on the adequate dependence of their refractive indices on the amplitude of 
light fields.
It is well known that the figure of merit of a suitable material for practical 
devices includes a fast and strong response to the light field as well as low 
losses \cite{stegeman97} which has motivated 
an active search for optical materials with the appropriate properties 
\cite{nlomat}.

On the other hand, a significant breakthrough in Quantum Optics 
has been the realization of giant optical nonlinearities in gases by 
means of atomic coherence and interference \cite{Field91}. 
A technique that has attracted much attention is
electromagnetically induced transparency (EIT) \cite{Harris97,Rathe93,Szymanowski94}, 
in which an opaque medium  becomes 
transparent to a probe laser beam by the addition of an appropriate coupling laser beam.
The adequate choice of an atomic level scheme and driving fields can yield 
to controlable nonlinearities with very interesting applications in the 
design of nonlinear optical devices.
This has been the basis for many studies on the resonant enhancement of 
nonlinear optical phenomena via EIT \cite{Andre2005,
schmidt96-imamoglu97,Harris,Petrosyan,Fleischhauer,hong03,soliton2}. 
However, only a few of these works have investigated the formation of transverse 
solitons \cite{hong03} and mostly considering the role played by the 
giant Kerr nonlinearity. 

In this paper we study the optical properties of a system where atomic 
coherence can be used to control the dependence of the refractive index on the 
amplitude of the light field. Many novel nonlinear optical phenomena beyond 
the giant Kerr effect are described, the most interesting being 
the obtention of the so-called \emph{liquid light condensates}
\cite{michinel02}, i. e. robust solitonic distributions of light with analogies
to ordinary fluid dropplets. 

%%%%%%%%%%%%%%%%%%%%%%%%%%%% FIG 1 %%%%%%%%%%%%%%%%%%%%%%%%%%%%%%%%%%%%%%%%%%
\begin{figure}
{\centering \resizebox*{1\columnwidth}{!}{\includegraphics{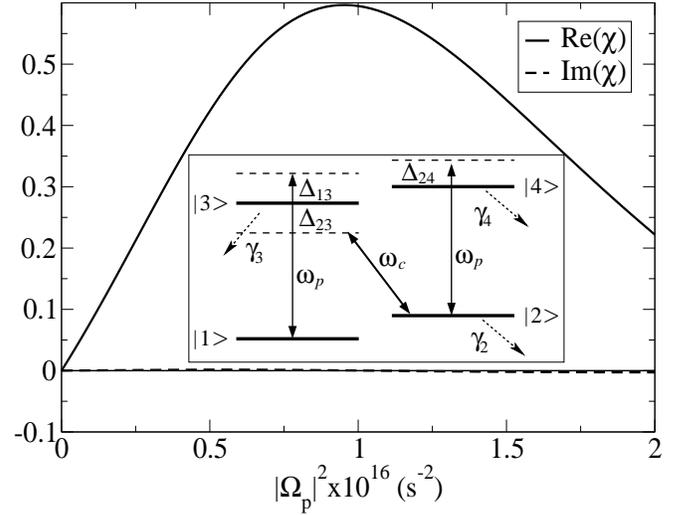}} \par}
\caption{
Real (solid curve) and imaginary (dashed curve) parts of the susceptibility 
$\chi$ given by Eq. (\ref{suscep}) for $\gamma=30$ MHz, 
$\gamma_{2}=10^{-8}\gamma$,$\gamma_3=\gamma_4=0.006\gamma$, 
$\Delta_{13}=\Delta_{23}=\gamma$, $\Delta_{24}=-1.5\gamma$,  
$\Omega_{c}=\gamma/2$, as a function of 
$|\Omega_{p}|^{2}=|\mu|^{2}|E_{p}|^{2}/4\hbar^{2}$.
{\em Inset}: Schematic plot of the energy levels and optical couplings of 
the four-level atomic system.  
\label{suscesq}}
\end{figure}
%%%%%%%%%%%%%%%%%%%%%%%%%%%%%%%%%%%%%%%%%%%%%%%%%%%%%%%%%%%%%%%%%%%%%%%%%%%%%

%-------------------- PHYSICAL MODEL ---------------------------------------

We consider the propagation of a weak probe light field of frequency 
$\omega_p$ in a medium composed of four-level atoms and a 
coupling light field of frequency $\omega_c$ (see e.g. \cite{hong03}).
A scheme of our system is shown in the 
inset of Fig. \ref{suscesq}.
In this kind of system, a coupling field of frequency $\omega_c$ changes 
the level structure \cite{autler-townes} 
and induces transparency for a probe beam of frequency $\omega_p$. 
A second effect is the enhancement of the optical Kerr nonlinearity.
 $\gamma_{2}$, 
$\gamma_{3}$, $\gamma_{4}$ denote the decay rates of the atomic states 
and $\Delta_{13}$, $\Delta_{23}$, $\Delta_{24}$ are light detunings. 
Direct electric-dipole transitions between the two ground states 
$\left|1\right\rangle$ and $\left|2\right\rangle$ are forbidden.

Paraxial propagation along $z$ of a probe laser beam through an optical 
medium, is given by:
\begin{equation}
\label{NLSE}
2ik_p\frac{\partial E_p}{\partial z}+ \left(\frac{\partial ^{2}}{\partial x ^{2}} 
+ \frac{\partial ^{2}}{\partial y ^{2}}\right) E_p=-k_p^2 \chi E_p.
\end{equation}
$k_p=2\pi/\lambda_p$ and $E_p$ are wave number and amplitude of the beam. 
The optical susceptibility $\chi$ in the  rotating wave and adiabatic 
approximations for EIT in the presence of a coupling beam $E_c$ 
takes the form \cite{hong03}:
\begin{multline}
\label{suscep}
\chi \left(E_{p},E_{c}\right) =-\frac{\eta|\mu|^{2}}{\epsilon_{0}\hbar\Gamma_{3}}
+\frac{\eta|\mu|^{2}}{\epsilon_{0}\hbar\Gamma_{3}A}\left(\frac{|\Omega_{c}|^{2}}{\Gamma_{3}}+\frac{|\Omega_{c}|^{2}|\Omega_{p}|^{2}}{\Gamma_{3}^{2}B}\right)\\
-\frac{\eta|\mu|^{2}|\Omega_{c}|^{2}|\Omega_{p}|^{2}}{\epsilon_{0}\hbar|\Gamma_{3}|^{2}\Gamma_{4}|A|^{2}}\left|1+\frac{|\Omega_{p}|^{2}}{\Gamma_{3}B}\right|^{2},
\end{multline}
where $|\Omega_{p,c}|^{2}=|\mu|^{2}|E_{p,c}|^{2}/4\hbar^{2}$ are 
squared Rabi frequencies and $\Gamma_{2}=\Delta_{23}-\Delta_{13}-i\gamma_{2}$, 
$\Gamma_{3}=\Delta_{13}+i\gamma_{3}$,
$\Gamma_{4}=\Delta_{24}+\Delta_{13}-\Delta_{23}+i\gamma_{4}$ and
$A=B+|\Omega_{c}|^{2}|\Omega_{p}|^{2}/(\Gamma_{3}^{2}B)$, with 
$B=\Gamma_{2}+|\Omega_{p}|^{2}/\Gamma_{4}+|\Omega_{c}|^{2}/\Gamma_{3}$,
being $\gamma_{2}$, $\gamma_{3}$ and $\gamma_{4}$ the decay rates of the 
atomic states and $\Delta_{13}$, $\Delta_{23}$ and $\Delta_{24}$ the light 
detunings.
For our numerical examples to be presented later we choose an atomic density $\eta=10^{14}$ cm$^{-3}$,
an electric dipole moment $\mu=3\times10^{-29}$ Cm (for alkalii atoms such 
as Rb or Ce, assuming for simplicity  $\mu_{13}=\mu_{23}=\mu_{24}=\mu$), $\gamma_{2}=10^{-8}\gamma$, $\gamma_3=\gamma_4=0.006\gamma$ (with 
$\gamma=30$ MHz), $\Delta_{13}=\Delta_{23}=\gamma$, 
$\Delta_{24}=-1.5\gamma$, $\Omega_{c}=\gamma/2$ and $\lambda_p = $ 800 nm.

From Eq. (\ref{suscesq}) we get the
coefficients  $\chi^{(j)}$ of the Taylor
expansion of the susceptibility 
$\chi = \sum^{\infty}_{j=0} \chi^{(2j+1)} \left| E_p\right|^{2j}$:
\begin{subequations}
\label{chito}
\begin{eqnarray}
\chi^{(1)} & = &-\eta|\mu|^{2}\Gamma_2/(\epsilon_{0}\hbar C) \label{chi1} \\
\chi^{(3)} & = & \frac{\eta|\mu|^{4}|\Omega_{c}|^{2}}{4\epsilon_{0}\hbar^{3}}\left[\frac{\frac{1}{\Gamma_{3}}-\frac{1}{\Gamma_{4}}}{C^{2}}
-\frac{|\Omega_{c}|^{2}}{\Gamma_{3}C^{3}}-\frac{1}{\Gamma_{4}|C|^{2}}\right] \label{chi3} \\
\chi^{(5)}& = & \frac{\eta|\mu|^6|\Omega_{c}|^{2}}{16\epsilon_{0}\hbar^5\Gamma_4}\left[\frac{D}{C^3}+\frac{|\Omega_c|^2(3-\Gamma_4/\Gamma_{3})}{C^{4}}\right.\nonumber\\
& & \left.+\frac{|\Omega_{c}|^{4}\Gamma_4}{C^{5}\Gamma_{3}}+\frac{D+\frac{|\Omega_{c}|^{2}}{C}}{C|C|^2}+\frac{D^{\ast}+\frac{|\Omega_c|^2}{C^{\ast}}}{C^{\ast}|C|^2}
\right] \label{(chi5)}
\end{eqnarray}
\end{subequations}
 $C=\Gamma_{2}\Gamma_{3}+|\Omega_{c}|^{2}$, $D=\Gamma_{3}/\Gamma_{4}-1$.
%%%%%%%%%%%%%%%%%%%%%%%%%%%% FIG 2 %%%%%%%%%%%%%%%%%%%%%%%%%%%%%%%%%%%%%%%%%%
\begin{figure}
{\centering \resizebox*{1\columnwidth}{!}{\includegraphics{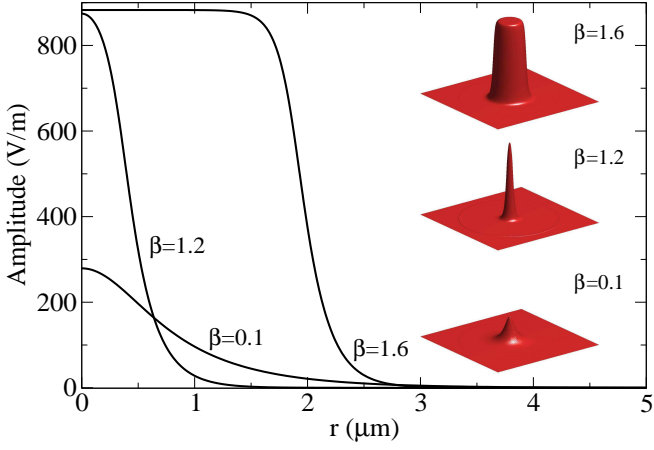}} \par}
\caption{(Color online).
Solutions of Eq. (\ref{eig}) for $\beta = 0.1 \ \mu$m$^{-1}$, 1.2 $\mu$m$^{-1}$ 
and 1.6 $\mu$m$^{-1}$ for our parameter set (values given in the text).
\label{eigenst}}
\end{figure}
%%%%%%%%%%%%%%%%%%%%%%%%%%%%%%%%%%%%%%%%%%%%%%%%%%%%%%%%%%%%%%%%%%%%%%%%%%%%%

In Fig. \ref{suscesq} we show the real and imaginary parts of the 
susceptibility $\chi$ as a function of the squared Rabi frequency of 
the probe light $|\Omega_{p}|^{2}=|\mu|^{2}|E_{p}|^{2}/4\hbar^{2}$, for our parameter choice. As it can be appreciated from  $\text{Re}\left(\chi(\Omega_p)\right)$ 
(see Fig. \ref{suscesq}),  the real part of the susceptibility of the medium 
grows linearly with $|E_p|^2$ for low powers (due to the effect of a positive 
$\chi^{(3)}_R$) and decreases for high powers (due to a 
negative $\chi^{(5)}_R$), while the losses are comparatively small in this range. Thus, we have a balance of diffraction plus self-focusing for low field amplitudes and self-defocusing for larger amplitudes. This type of competition is also found in media with the so-called nonlinearity of cubic-quintic type, i.e. those with  a refractive  index of the form $n=n_0+n_2|E|^2+n_4|E|^4$. These nonlinearities 
have attracted a lot of theoretical attention recently \cite{michinel02,piekara97,vortexa,vortexb,vortices-st,clusters} because of their predicted ability, when $n_4<0$,  to prevent collapse of laser beams for sufficiently large powers, thus  
yielding to different stable two-dimensional light distributions \cite{piekara97}. 
The robustness of these \emph{light bullets}  has been recently connected with the formation of
a liquid light condensate with surface tension properties similar to those of usual liquids \cite{michinel02}.
These media are able to support stable vortex beams \cite{vortexa,vortexb,vortices-st} and display interesting nonlinear phenomena  \cite{clusters}. 

For our choice of parameters, using Eqs. (\ref{chito}) 
%$\chi^{(1)}_R=-8.1817\cdot10^{-16}$, $\chi^{(1)}_I=2.0454\cdot10^{-8}$,
%$\chi^{(3)}_R=1.5529\cdot10^{-6}$m$^2/$V$^2$, 
%$\chi^{(3)}_I=6.2117\cdot10^{-9}$m$^2/$V$^2$,
%$\chi^{(5)}_R=-1.8868\cdot10^{-16}$m$^4/$V$^4$, and
%$\chi^{(5)}_I=1.1789\cdot10^{-14}$m$^4/$V$^4$. 
and the relation $n(E_p)\simeq n_{0}+(\chi^{(3)}_R/2n_{0})\left| E_p\right| ^{2}+\frac{1}{2n_{0}}\left[\chi^{(5)}_R-(\chi^{(3)}_R/2n_{0})^{2}\right] \left| E_p\right|^{4}+\cdots,$ we obtain $n_{2}^R=7.7646\cdot10^{-7}$m$^2/$V$^2$, $n_4^R=-3.0154\cdot10^{-13}$m$^4/$V$^4$, which are, respectively, $\sim 10^{13}$ 
and $\sim 10^{22}$ larger than those measured for usual nonlinear optical 
materials \cite{saltiel97}. These facts provide some analogies between our system and CQ media. However, the contribution of higher order and dissipative terms will be relevant for us.

%%%%%%%%%%%%%%%%%%%%%%%%%%%% FIG 3 %%%%%%%%%%%%%%%%%%%%%%%%%%%%%%%%%%%%%%%%%%
\begin{figure}
{\centering \resizebox*{1\columnwidth}{!}{\includegraphics{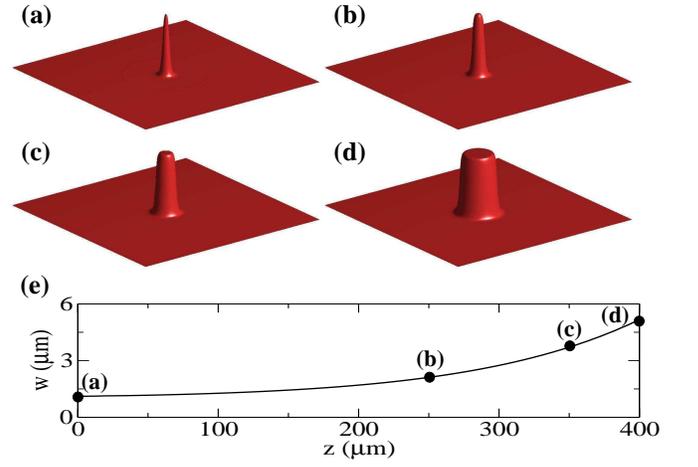}} \par}
\caption{(Color online).
Propagation of an eigenstate of Eq. (4) with  
$\beta=1.3\ \mu$m$^{-1}$ in as medium with the full complex susceptibility
given by Eq. (\ref{suscep}). (a-d) 3D plots of $|E_p(x,y,z)|^2$ 
for $z=0, 250, 350$ and $450$ $\mu$m. (e) Beam width $w(z)$.
\label{b1_3}}
\end{figure}
%%%%%%%%%%%%%%%%%%%%%%%%%%%%%%%%%%%%%%%%%%%%%%%%%%%%%%%%%%%%%%%%%%%%%%%%%%%%%

%-------------- STATIONARY STATES ----------------------------------------
 First we will construct 
stationary transverse self-trapped solutions of Eq. (\ref{NLSE}) of the form: 
$E_p(r,z)=\psi_{\ell}(r)e^{i\beta z}e^{i\ell \theta}$, 
where $\beta$ is the propagation constant. For $\ell \neq 0$ the beam 
host a vortex of topological charge $\ell$. To this end we set $\chi_{I} = 0$ 
and solve numerically the problem:
\begin{equation}\label{eig}
\left[\frac{d^2}{dr^2} + \frac{1}{r}\frac{d}{dr} - \frac{\ell^2}{r^2} + k_p^2 \chi_R(\psi_{\ell}) - 2 k_p\beta\right] \psi_{\ell} = 0,
\end{equation}
with  boundary conditions $\psi_{\ell}'(0) = 0$ and $\psi_{\ell}(\infty) = 0$.
 
This gives us stationary beam shapes corresponding to different powers as a 
function of $\beta$. Let us first consider beams with $\ell = 0$. 
In Fig. \ref{eigenst} we show the results for $\beta = 0.1$  $\mu$m$^{-1}$, 
1.2 $\mu$m$^{-1}$ and 1.6 $\mu$m$^{-1}$. Low values of $\beta$ yield to 
light distributions with quasi-Gaussian profiles. As $\beta$ is incremented, 
the spatial shapes become narrower, but still keeping a Gaussian shape. 
For larger values of $\beta$, the beam flux grows rapidly and the peak 
intensity of the light distribution saturates due to the effect of a 
negative $n_4$, yielding to light distributions with almost super-Gaussian 
profiles. Due to the giant nonlinear response, these powers can be 
experimentally achieved by using mW continuous lasers provided the sources 
are highly stabilized in frequency (typically $1$ MHz bandwidth). A warning is in order: 
for our parameter combination, the probe beam has a power smaller but close to that of the coupling beam thus a fully quantitative treatment should consider a vector extension of Eqs (\ref{NLSE}) 
including  the effect of the probe beam on the coupling beam. Since this extension makes the analysis even more complex in this paper we restrict ourselves to the scalar model and the full vector model will be the subject of future research.

%%%%%%%%%%%%%%%%%%%%%%%%%%%%%% FIG 4 %%%%%%%%%%%%%%%%%%%%%%%%%%%%%%%%%%%%%%%
\begin{figure}
{\centering \resizebox*{1\columnwidth}{!}{\includegraphics{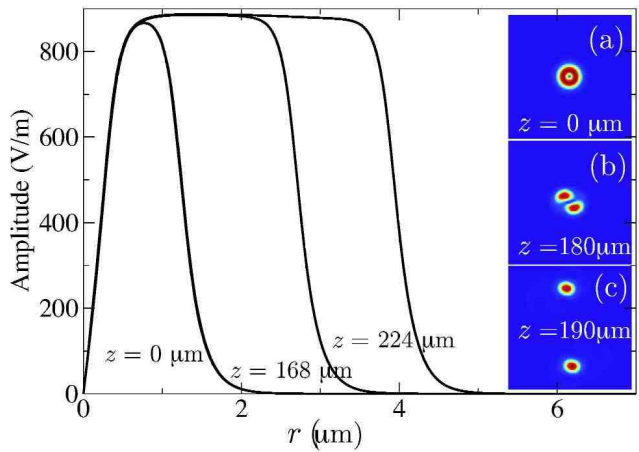}} \par}
\caption{(Color online).
Evolution of two different eigenstates with $\ell = 1$ for several 
propagation distances (in $\mu$m).For $\beta = 1.5\mu$m$^{-1}$, the gain 
enlarges the radial size of the beam preserving the vortex (solid lines).
Insets (a-c): Evolution of the eigenstate with $\beta=1.3 \mu$m$^{-1}$ for 
$z=0, 180$ and  $190\mu m$.
\label{vorti}}
\end{figure}
%%%%%%%%%%%%%%%%%%%%%%%%%%%%%%%%%%%%%%%%%%%%%%%%%%%%%%%%%%%%%%%%%%%%%%%%%%%

%%%%%%%%%%%%%%%%%%%%%%%%%%%% FIG 5 %%%%%%%%%%%%%%%%%%%%%%%%%%%%%%%%%%%%%%%%%%
\begin{figure}
{\centering \resizebox*{1\columnwidth}{!}{\includegraphics{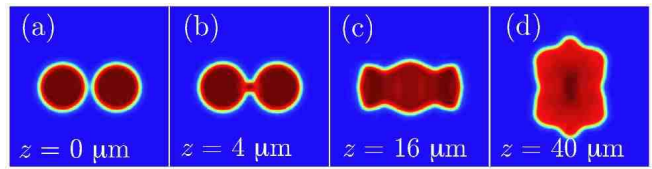}} \par}
\caption{(Color online).
Coalescence of two eigenstates with $\beta=1.6\ \mu$m$^{-1}$ 
launched in  parallel with the same phases. 
\label{coalescence}}
\end{figure}
%%%%%%%%%%%%%%%%%%%%%%%%%%%%%%%%%%%%%%%%%%%%%%%%%%%%%%%%%%%%%%%%%%%%%%%%%%%%%

We have used the eigenstates of the nondissipative case as input conditions 
for propagation in a medium with the full complex susceptibility of 
Eq.(\ref{suscep}) (i.e. 
including the imaginary part of $\chi$). 
We observe that for $\beta=1.2\  \mu$m$^{-1}$ the eigenstate keeps its shape 
while propagating in such a medium. In this situation the amount of energy 
pumped into the soliton and taken out by the linear and nonlinear gain and 
dissipation respectively achieves an equilibrium \cite{solidisi}. 
Eigenstates with $\beta<1.2\ \mu$m$^{-1}$ tend to spread during propagation 
since they do not achieve the critical power for the formation of a soliton,
while those with $\beta>1.2\ \mu$m$^{-1}$ 
keep their peak amplitude and increase their width during propagation as it 
can be seen in Fig. \ref{b1_3}. In this situation the energy available in 
the medium is the responsible for the broadening of the liquid light droplet 
in a similar way to the process of growth of a fluid droplet in a 
supersaturated atmosphere. This means that although small, nonconservative 
effects play an important role in the propagation of wavepackets for this 
set of parameters. We have found numerically with a very high accuracy that 
the radius of the light droplet varies as $R(z) \sim z^3$, which is 
faster than typical fluid droplet growth phenomena \cite{liquids} or 
models similar to C-Q ones such as the Ginzburg-Landau equations \cite{Maxi}. 
In both cases the growth takes the form $z^q$ with $q<1$. 
Even pure diffractive propagation leads an exponent $q=1$, which is smaller 
than the one observed in our system.

Next, we have constructed eigenstates of Eq. (\ref{eig}) with  $\ell  = 1$. 
Eigenstates with $\beta \leq 1.3 \ \mu$m are unstable under propagation 
due to the presence of gain and the vortex breaks into two fundamental beams 
as shown in Fig. \ref{vorti}(a-c). However, for $\beta > 1.3 \ \mu$m 
the beams reach a critical value of the energy \cite{vortexa} so that the liquid 
light condensate is formed and the surface tension is able to sustain the vortex 
within the beam. Thus, the effect of gain is to enlarge the radial size of 
the beam without destroying the vortex.
As the beam propagates the width of the beam surrounding the vortex increases 
while keeping the peak density constant, which again resembles the growth of 
an incompressible fluid. It is remarkable that the maximum densities in 
Figs. \ref{eigenst} and \ref{vorti} are very similar.

\begin{figure}
{\centering \resizebox*{1\columnwidth}{!}{\includegraphics{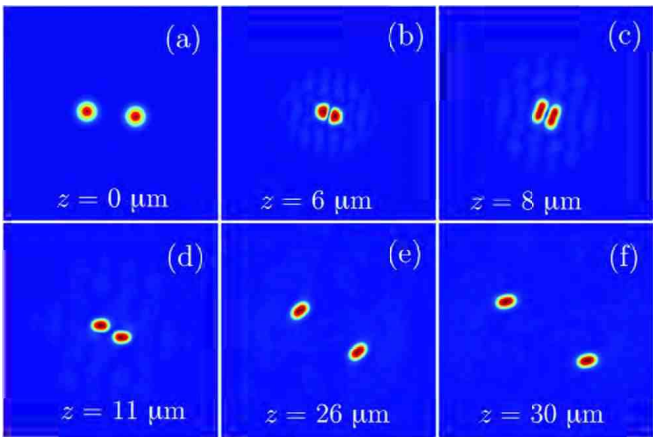}} \par}
\caption{(Color online).
Collision of two hipergaussian beams launched with opposite phases, 
amplitudes $A=890$ V/m and width 
$w=4\ \mu$m, separated $4\ \mu$m and being one of them slightly 
displaced in the y-axis. The initial angle is $0.25$ rad. 
\label{collision}}
\end{figure}

 To study the robustness of these solitons we have made a series of numerical experiments
with collisions of different beams (we show results for $\ell =0$). First we have launched initially parallel beams in phase 
corresponding to eigenstates with  $\beta=1.6 \ \mu$m$^{-1}$. Their mutual effective interaction, as it happens with solitons of the 1D nonlinear Schr\"odinger equation, is attractive and leads to 
their fusion and subsequent trasverse oscillations of the new bound state
as it can be seen in Fig. \ref{coalescence}.
In a different series of numerical experiments we have studied the collisions of the same beams launched initially with opposite phases as shown in Fig. \ref{collision}.  As it can be seen, both beams survive the collision behaving as light droplets, although slight transverse oscillations of the beams are observed after the collision by excitation of surface modes. Both phenomena arise in fluid droplets collisions \cite{Melissa}.

%%%%%%%%%%%%%%%%%%%%%%%%%%%% FIG 6 %%%%%%%%%%%%%%%%%%%%%%%%%%%%%%%%%%%%%%%%%%
%%%%%%%%%%%%%%%%%%%%%%%%%%%%%%%%%%%%%%%%%%%%%%%%%%%%%%%%%%%%%%%%%%%%%%%%%%%%%

\begin{figure}
{\centering \resizebox*{1\columnwidth}{!}{\includegraphics{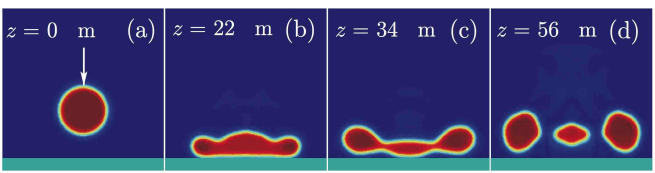}} \par}
\caption{(Color online).
Total reflection at a nonlinear-linear interface of an eigenstate 
with $\beta=1.6\ \mu$m$^{-1}$ and an incidence angle of $0.017$ rad, at 
$z=0, 22, 34$ and $56 \ \mu$m. 
\label{reflection}}
\end{figure}

Finally, we have launched an eigenstate with $\beta=1.6\ \mu$m$^{-1}$ with an incidence 
angle of $0.017$ rad against the frontier between a nonlinear and 
a linear material. The results are shown in Fig. \ref{reflection}, where the breakup of the beam into smaller 
droplets is observed analogously to splashing fluid droplets \cite{splash}.

%%%%%%%%%%%%%%%%%%%%%%%%%%%% FIG 7 %%%%%%%%%%%%%%%%%%%%%%%%%%%%%%%%%%%%%%%%%%
%%%%%%%%%%%%%%%%%%%%%%%%%%%%%%%%%%%%%%%%%%%%%%%%%%%%%%%%%%%%%%%%%%%%%%%%%%%%%

%---------------------   CONCLUSIONS  -------------------------------------

In conclusion,  we have shown that the adequate choice of the parameters of a 
specific EIT scheme leads to a giant response for both $n_{2}$
and $n_{4}$ (but different signs) and could allow to obtain stable two-dimensional liquid light
condensates with  surface tension properties similar to those of
usual liquids. Our theoretical and computational results could be the basis for real experiments in 
nonlinear optics with continuous mW lasers showing this phase transition and nice liquid-like 
properties of light.

%---------------------   ACKNOWLEDGMENTS  ---------------------------------
 This work has been supported by grants  FIS2004-02466, BFM2003-02832 (Ministerio de Educaci\'on y 
Ciencia),  PGIDIT04TIC383001PR (Xunta de Galicia) and PAI-05-001 (Consejer\'{\i}a de Educaci\'on y Ciencia, Junta de Comunidades de Castilla-La Mancha). The authors want to acknowledge 
M. Fleischhauer, R. Corbal\'an and J. Mompart for discussions.

%-----------------------------------------------------------------------------

\end{document}